\documentclass[12pt,preprint]{aastex}

\slugcomment{}

\shorttitle{L483}
\shortauthors{Connelley et al.}

\begin{document}


\title{A Photometrically and Morphologically Variable Infrared Nebula in L483}


\author{Michael S. Connelley \altaffilmark{1,2}}
\affil{NASA Ames Research Center}

\author{Klaus W. Hodapp \altaffilmark{3}}
\affil{University of Hawai$'$i}

\and

\author{Gary A. Fuller \altaffilmark{4}}
\affil{University of Manchester}


\altaffiltext{1}{NASA Ames Research Center, MS 245-6, Moffett Field, CA 94035}
\altaffiltext{2}{Visiting Astronomer at the Infrared Telescope Facility, which is operated by the University of Hawaii under Cooperative Agreement no. NCC 5-538 with the National Aeronautics and Space Administration, Science Mission Directorate, Planetary Astronomy Program.}
\altaffiltext{3}{University of Hawai$'$i Institute for Astronomy, 640 N. Aohoku Pl., Hilo HI 96720}
\altaffiltext{4}{Jodrell Bank Centre for Astrophysics, Alan Turing Building, University of Manchester, Manchester, M13 9PL, UK}


\begin{abstract}

   We present narrow and broad K-band observations of the Class 0/I source IRAS 18148-0440 that span 17 years.  The infrared nebula associated with this protostar in the L483 dark cloud is both morphologically and photometrically variable on a time scale of only a few months.  This nebula appears to be an infrared analogue to other well-known optically visible variable nebulae associated with young stars, such as Hubble's Variable Nebula.  Along with Cepheus A, this is one of the first large variable nebulae to be found that is only visible in the infrared.  The variability of this nebula is most likely due to changing illumination of the cloud rather than any motion of the structure in the nebula.  Both morphological and photometric changes are observed on a time scale only a few times longer than the light crossing time of the nebula, suggesting very rapid intrinsic changes in the illumination of the nebula.  Our narrow-band observations also found that H$_{2}$ knots are found nearly twice as far to the east of the source as to its west, and that H$_{2}$ emission extends farther east of the source than the previously known CO outflow.

\end{abstract}


%% subject header.  Objects should be in the appropriate "individual"
%% headers (e.g. quasars: individual, stars: individual, etc.) with the
%% additional provision that the total number of headers, including each
%% individual object, not exceed six.  The \objectname{} macro, and its
%% alias \object{}, is used to mark each object.  The macro takes the object
%% name as its primary argument.  This name will appear in the paper
%% and serve as the link's anchor in the electronic edition if the name
%% is recognized by the data centers.  The macro also takes an optional
%% argument in parentheses in cases where the data center identification
%% differs from what is to be printed in the paper.


%% chosen the first three characters of the first author's name plusF

\section{Introduction}

   Variable nebulae, well-known examples being Hind's Variable Nebula, R CrA, Hubble's Variable Nebula, and PV Cephei, are now known to be associated with embedded young stars and were discovered through visible light observations.  Hind's Variable Nebula (Hind 1864), discovered near T Tauri in 1852, was easily visible in 1855-56, but was not seen by Struve in 1861 \citep{Bar1895}.  However, Struve observed a small nebula 4$'$ to the west that had not been previously visible.  The nebula around R CrA was found to be variable before 1890.  \citet{Kno1916} describes the changes in the morphology of the nebula, how the changes in the brightness of the nebula are uncorrelated with the brightness of R CrA itself, and that changes are noticeable on a time scale of a week but are not noticeable between consecutive nights.  \citet{Hub1916} found that several features in NGC 2261 (Hubble's Variable Nebula) were observed to vanish, then later reappear in the same location.  \citet{Lam1926} successfully explained the changes in morphology as changes in the illumination of the cloud rather than on account of any actual motion of the structure in the nebula.  PV Cephei is another young star with an optically visible variable nebula (Cohen et al. 1977; Gyul$'$budagyan \& Magakyan, 1977).  Morphological changes are seen on photographic plates within a month, and the point-like source at the southern tip of the nebula varied by over four magnitudes in less than six months.  \citet{Hod2008} found that the infrared reflection nebula illuminated by Cep A also shows photometric and morphological variability.  Changes in the illumination of the circumstellar material were proposed to account for the observed changes in morphology.  McNeil's Nebula \citep{Rei2004} is a recent example of a large, optically visible variable reflection nebula that became visible when V1647 Ori experienced an eruption.  Optical photometry \citep{Bri2004} shows that the photometry of the nebula closely followed the photometry of the central star, and that the morphology of the nebula \citep{Fed2007} was largely static during the eruption.

  IRAS 18148-0440 was identified as a candidate protostar by \citet{Par1988} on account of its IRAS colors and coincidence with the L483 dark cloud.  This source has been classified as a Class 0 YSO by \citet{And2000}, and as a Class I YSO by \citet{Pez2002} based on its ISO 60~$\mu$m~$-$~100~$\mu$m vs. 100~$\mu$m~$-$~170~$\mu$m colors.  \citet{Taf2000} suggest that this source is in transition between Class 0 and I on account that it has an SED characteristic of a Class 0 source but its outflow properties and near-IR reflection nebula suggest that it is a Class I source.  Compared to Cep A, the source in L483 is a lower mass, lower luminosity object that is likely to be at an earlier evolutionary state.  The distance to L483 has been given as 200~pc by \citet{Dam1985} and \citet{Goo1993}, and 250~pc by \citet{Fel1992}. 

   Our observations have found that IRAS 18148-0440 is the source of a new variable nebula that is only visible in the infrared.   \citet{Ful1995} presented images of the infrared nebula at J, H, K, and H$_{2}$, demonstrating the alignment of the H$_{2}$ knots with the CO outflow and the coincidence of the near-IR reflection nebula with the blue shifted lobe of the outflow.  They estimate that the dynamical age of the outflow is 1.3$\times10^{4}$ years, and that the outflow is tilted $40^{\circ}$ to the line of sight.  \citet{Buc1999} studied the excitation temperature and velocity of the shocks associated with the outflow using long slit spectroscopic observations of the H$_{2}$ emission from knots in the jet.  Derived properties of L483 are summarized by \citet{Ful2000}, who observed submillimeter dust continuum emission and NH$_{3}$ line emission.  They interpret the NH$_{3}$ line widths and velocities as a signature of infall, and present a model to account for their observations.

   In this paper we present K-band images of IRAS 18148-0440 collected over the span of 17 years, as well as an image that shows the full extent and asymmetry of the H$_{2}$ knots.  The observations show that the infrared nebula is both photometrically and morphologically variable.  Our goal is to present a time series of K-band images and photometry of this nebula to demonstrate its photometric and morphological variability.

\section{Observations and Photometry}
   We observed IRAS 18148-0440 (18$^{h}$ 17$^{m}$ 29.8$^{s}$ $-$04$^{\circ}$ 39\arcmin ~38\arcsec ~J2000) in L483 on 9 occasions, and include observations from \citet{Ful1995} and 2MASS \citep{Skr2006} for a total of 11 epochs.  The location of the centimeter source in the nebula given by \citet{Ful2000} is less than an arcsecond from the position noted above.  The observations used in this paper are summarized in Table 1.  The data acquisition and reduction for the data taken in Oct. 1990 and June 1993 are described in \citet{Hod1994} and \citet{Ful1995}, respectively.  Observations by \citet{Ful1995} were taken through Barr near-infrared filters.  The data taken with QUIRC \citep{Hod1996} at K and K$'$ \citep{Wai1992} and the SpeX guider \citep{Ray2003} are in the MKO photometric system (Simons \& Tokunaga 2002, Tokunaga \& Simons 2002).
   
     All data from QUIRC and SpeX were taken while dithering, usually with an equal number of dithered sky frames offset from the target.  For the data reduction, a dark frame, of the same exposure time as the data, was subtracted from each frame.  A sky flat was constructed from the dithered images of the sky near the target by first scaling these images to the same median value, then median combining them together with a min-max rejection.  We divided each image of the target by a normalized sky flat, then subtracted the median sky level from each image.  The resulting images were then aligned and averaged together with a sigma clipping rejection.  

   The observations of L483 through the H$_{2}$ v=1-0 S(1) 2.121~$\mu$m filter have a slightly different plate scale than the K-band images.  Before the K-band image could be subtracted from the H$_{2}$ image, we used the IRAF tools geotrans and geomap to match the star fields.  We then used a Gaussian convolution to blur the K-band image to match the FWHM of the H$_{2}$ image, scaled the resulting image so that the stars had the same peak counts, then subtracted the scaled K-band image from the H$_{2}$ image.  
      
   We measured the photometry of the L483 infrared nebula relative to four nearby field stars since some of the nights were non-photometric.  The photometry of the field stars was measured using an aperture 1.5 times the FWHM to ensure that the same fraction of the stars' light was in the aperture despite the widely varying angular resolution of our data set.  Since the nebula is spatially extended, and elongated east to west, we used a rectangular aperture measuring 40\arcsec~ east to west and 20\arcsec~ north to south to measure the photometry of the nebula.  The average of the counts from the nearby sky was subtracted from the image before the counts in the rectangular aperture were summed.  

\section{Discussion}

\subsection{Classification} 

   Class 0 YSOs are characterized by having an SED that is a cool single temperature blackbody \citep{Lad1991}, with little or no flux in the near-IR, whereas the SED of a Class I YSO can not be fit by a single temperature blackbody and has significant flux in the near-IR.  As mentioned in the introduction, various authors have classified this object as a Class 0 YSO, a Class I YSO, or an object in transition between the two.  We combined data from 2MASS (J, H, and K), Spitzer (3.6~$\mu$m, 4.5~$\mu$m, 5.8~$\mu$m, 8.0~$\mu$m), IRAS (12~$\mu$m, 25~$\mu$m, 60~$\mu$m, 100~$\mu$m) and JCMT/SCUBA (850~$\mu$m, Fuller \& Wooten 2000) to create the SED shown in Figure 1.  We note that the 12~$\mu$m flux is only an upper limit.  

   We chose to fit the IRAS fluxes with a blackbody function since that wavelength range is least affected by flux from the cold envelope and is least attenuated by extinction.  When these data are fit with a 45~K blackbody (the best-fit temperature), the 850~$\mu$m flux measurement is much greater than predicted by the fit.  If the 850~$\mu$m flux measurement is included, the best-fit temperature is 35~K.  The 850~$\mu$m map presented by \citet{Ful2000} shows emission extending $\sim$60\arcsec~ from the source.  As such, much of the emission at this wavelength is likely to be from the cool envelope around the source rather than the source itself.  Despite the large amount of extinction towards this object, there is still significant amount of flux at wavelengths shorter than 10~$\mu$m.  We did not fit a separate blackbody function to the flux measurements at wavelengths shorter than 10~$\mu$m since they are strongly attenuated by extinction and also strongly affected by scattered light.  This excess flux at wavelengths shorter than 10~$\mu$m supports the conclusion that L483 is not a Class 0 YSO, but has more in common with deeply embedded Class I YSOs.  

\subsection{Change in Brightness} 

    Accurate relative photometry of this object is very difficult.  First, the object is large, diffuse, and irregularly shaped.  While there is a dearth of field stars immediately around the nebula due to the dark cloud that embeds the source, the nearby field stars can affect the photometry if they lie either in the aperture or in the region used to fit the sky level.  Second, the observations were not part of a focused effort, and were made over 17 years using 5 different cameras and filters from at least 3 different photometric systems.  However, the different filters and color responses of the cameras affect our relative photometry of the nebula only so far as the color of the nebula is different than the color of the field stars used as the photometric references.  The mean 2MASS H$-$K color of the reference stars we used is 1.66, which closely matches the 2MASS color of the nebula (H$-$K=1.85).  Our H and K observations in July 1999 show that the nebula has a H$-$K color of 2.06, similar to the 2MASS color.
    
      The symbols in Figure 2 represent relative photometry of the whole nebula using different field stars.   In some cases, only one star was within the field of view that could be used as a photometric reference.  Since the relative photometry of the nebula from different field stars are very close to each other, the color effects described in the previous paragraph do not appear to significantly affect our photometry compared to the photometric variability of the nebula.  Furthermore, using the observed colors to transform the photometry to a common photometric system can be misleading since the color of the nebula may also be variable.  \citet{Ful1995} state that the K-band peak of the nebula has an H$-$K color of 4.1, whereas the H$-$K color of the brightest part of the nebula in our July 1999 data is 2.30 in an 8\arcsec~ aperture.  Variability of the color of the nebula may be due to a combination of the color gradient across the nebula (the nebula is bluer towards the west) with the changes in the brightness of various knots in the nebula.  With our current data set, we cannot determine if individual knots have a variable color.

   Over the 17-year span of observations, the brightness of the nebula changed by over two magnitudes, and changes were observed on a time scale comparable to the light crossing time of the nebula. The integrated K-band magnitude of the nebula varied from 9.8 to 12.0 (Figure 2), and changes in brightness by about 1 magnitude were observed within a two-month time span.  It is likely that significant changes in the brightness of the nebula occur on much shorter time scales than the spans between our observations.  We did not observe frequently enough to determine the minimum time scale of the variability, or to determine if the variability is periodic.  The center of the nebula at K-band is $\sim15$\arcsec~ away from the source, or 3000~AU at a distance of 200~pc.  As such, the light crossing time is $\sim17$ days.  The illumination of the nebula must be significantly and rapidly changing to show significant brightness changes on a time scale of, at most, four times longer than the light crossing time.  Our photometry also shows that the nebula appears to be fading at an average rate of $\sim0.08$ magnitudes per year.  

\subsection{Morphological Variability} 

   We noticed changes in the apparent morphology of the nebula over the 17-year span of our observations.   Although we have images from \citet{Ful1995} from 1993, 2MASS from 1999, and our data from 2002, these are not included in Figures 3 and 4 since they do not have sufficient angular resolution to show changes in morphological details.   The 8 epochs of data that are used in the following analysis were added together to create a master image of the nebula that is representative of its average state.  Figure 3 labels several regions that were observed to change.  Figure 4 shows a K or K$'$ image of the nebula as well as the difference between the image taken on a particular date and the master image.  For each difference, the input image was scaled to match the brightness of the master image to compensate for the overall variability of the nebula and to better show the morphological changes. The angular resolution within these figures varies by over a factor of 2, and is documented in Table 1.  

  The differenced images in the right column of Figure 4 clearly show the morphological changes in the nebula.  In 1990, the central part of the western lobe of the nebula was brighter, whereas the opposite was true 9 years later.  In July 1999, the southern half of the western lobe was unusually bright, but two months later this had changed.  Region A was brighter than average in 2003 only.  By 2004, region B was unusually bright, which was also observed in 2007.  The excess illumination of the nebula in 2007 appears to point away from the expected location of the source (roughly between regions A and C).  

  The morphological features in the nebula tend to vary in brightness together, but can change brightness independently of each other.  Light curves for the four regions outlined in Figure 3 are shown in Figure 5.  Region A appears to brighten significantly from 1999 to 2003.  However, its integrated K-band magnitude brightens by only $\sim0.2$ magnitudes whereas the other three regions fade by $\sim1$ magnitude.  Region A is typically fainter than the other three regions, and became much fainter than the other regions between 2004 and 2007.  Regions B and D had a similar brightness until May 2004, when regions C and D faded by $\sim0.6$ magnitudes but region B stayed the same brightness.  Although regions B and D had a similar brightness in August 2004, region D was $\sim0.9$ magnitudes fainter ($\Delta$K=2 vs. $\Delta$K=0.7) than region B in July 2007.  The knot that is pointed out in Figure 3 is a feature that reappears in the same place after a period of not being visible.  It is faintly visible in our July 1999 observation, but it is much more prominent just three months later.  It is not visible in 2003, but it is again prominently visible in May 2004.  By July 2007, it is only faintly visible.  

  We note that in the course of our observations, neither the position nor the morphology of these knots, as well as the features around them, appear to change.  Rather, only their apparent brightness changes.  Similarly, \citet{Lam1926} noticed that morphological features in Hubble's Variable Nebula would disappear only to once again become visible in the same place.  This led to his conclusion that the most likely explanation for the observed changes in morphology is changes in the illumination of the nebula.  Since we observed very similar phenomena in IRAS 18148-0440, we propose that the cause of the morphological variability of the nebula are changes in the illumination of the nebula, possibly by opaque clouds very close to the source.  As noted above, morphological changes are noticeable within two months, with significant changes in the illumination of the cloud are observed to occur within a year (e.g. between July 2003 and May 2004).  If we assume that opaque clouds in Keplerian orbits around a 0.35~M$_{\odot}$ \citep{Ful2000} star are the cause of the change of illumination of the nebula, then the time scale of the observed variability corresponds the orbital period of the clouds.  In this case, a period of a year suggests that the distance from the central star to these clouds is of the order of $\sim1$~AU.  However, images taken several years apart show a similar pattern in the differenced images (e.g. Aug 2004 to Jul 2007), suggesting that the time scale for significant changes in the illumination of the cloud may be several years, in which case the obscuring clouds would be farther from the central star.

   Since this is one of a very few known infrared variable nebula, it is difficult to estimate how common such nebulae are.  Since there are only a few optically visible variable nebulae that are illuminated by T Tauri stars, and since there are an order of magnitude fewer Class I YSOs than T Tauri stars, infrared variable nebula may be very rare.  However, Class I YSOs are much more deeply embedded than T Tauri stars, and commonly have a reflection nebula visible in the near-IR \citep{Con2007}.  As such, future observations may show that there are variable near-IR nebulae around many other Class I YSOs.  

\subsection{H$_{2}$ Emission} 

  Figure 6 presents 2\arcmin~ wide images of IRAS 18148-0440 to illustrate the relationship between the near-IR reflection nebula, the H$_{2}$ knots, as well as the location of the source and the outflow cavity which are visible in the mid-IR.  The bright H$_{2}$ shock to the west, as well as the H$_{2}$ knots farthest east, are visible on account of thermal dust emission in Spitzer IRAC images at 3.6~$\mu$m, 4.5~$\mu$m, 5.8~$\mu$m, and 8.0~$\mu$m \citep{Eva2003}.  The infrared emission corresponding to the H$_{2}$ knots are found 65\arcsec~ to the west and 118\arcsec~ to the east of the source, the location of which is clearest in the 8.0~$\mu$m Spitzer image.  However, CO maps by \citet{Ful1995} and \citet{Taf2000} show that the CO outflow, which is aligned with the axis of the H$_{2}$ knots, extends about as far east of the source as to the west. Since the CO outflow extends only as far as the bright H$_{2}$ shock to the west, the eastern H$_{2}$ knots are farther east from the source than the CO emission contours shown in \citet{Taf2000}.  We note that the western knots appear brighter, likely because the western side of the nebula corresponds with the blue shifted lobe of the CO outflow.  The western knots are thus seen through less extinction than the H$_{2}$ knots from the more deeply embedded red shifted lobe of the outflow.  None of the H$_{2}$ knots correspond with conspicuous knots in the bright reflection nebula, and no H$_{2}$ knot is found within 20\arcsec~ of the source.

\subsection{Nebula Morphology} 

   \citet{Ful1995} showed that the near-IR reflection nebula is coincident and aligned with the blue shifted lobe of the CO outflow from the source.  The near-IR image of the nebula shows the half of a bipolar cavity that is closer to us, where flux from the central source is scattered by the inner walls of the bipolar cavity.  It is quite common for only one half of a (presumably) bipolar cavity to be visible in an infrared reflection nebula \citep{Con2007}.  Most, if not all, of the flux in the $8~\mu$m image is scattered light since the FWHM of the bright source in the center of the nebula is 4\farcs22 whereas the FWHM of a nearby star is 1\farcs91.  The centroid of the 24~$\mu$m source observed by Spitzer is 1\farcs1 to the west of the $8~\mu$m centroid.  At 24~$\mu$m, this source is observed to be unresolved since the FWHM of the 24~$\mu$m source is consistent with the FWHM of nearby field stars.  This is likely to be the true location of the protostar, and this location is shown in Figure 3.  

   The dark lane visible $\sim$2\farcs5 ($\sim$560 AU) west of the bright $8~\mu$m source appears to be coincident with the sharp eastern edge of the near-IR reflection nebula.  This dark lane appears to be similar to a disk shadow being projected into the cloud (Hodapp et al. 2004, Pontoppidan et al. 2005).  If this dark lane was a disk shadow, then the brightest part of the nebula should be on the closer (blue shifted) side of the bipolar cavity since forward scattering of light by dust is more efficient than back scattering, and light scattered on the closer side of the nebula suffers less extinction.  However, the bright $8~\mu$m source is on the farther (red shifted) side of the nebula, which is difficult to explain within the conventional model of a single source with a disk illuminating a bipolar cavity.  It is possible that the illuminating source of the nebula is totally obscured and the dark lane is a shadow of the disk being projected into the cloud, which would account for the dark lane being prominent from $1.6~\mu$m to $8~\mu$m.  In this case, the $8~\mu$m source may be scattered light from a nearby and less obscured companion.  

\section{Summary}

  We have presented K-band images of IRAS 18148-0440 in L483 spanning 17 years.  This Class0/I YSO is one of the first large nebulae that are only visible in the near infrared that has been found to be variable.  Morphological changes are noticeable within two months, and features are observed to disappear then reappear in the same location.  These changes are consistent with a changing illumination of the nebula, which may be caused by opaque clouds within $\sim1$~AU the protostar.  As such, this nebula is an infrared analogue to well known optically visible variable nebulae such as Hubble's Variable Nebula.  The overall brightness of the nebula appears to be slowly declining, with observed variations of a magnitude within two months and two magnitudes within a year.  Since these changes happen on a time scale a few times greater than the light crossing time of the nebula, the intrinsic changes in the illumination of the nebula must be rapid.  The H$_{2}$ shocks are visible in archival Spitzer images, and these shocks can be found nearly twice as far to the east of the source as to the west.  The SED of this source is consistent with it being a Class I YSO on account of its near-IR flux.  While Cep A \citep{Hod2008} and L483 host the first infrared variable nebulae identified to date, these are likely to be the first of many to be found.

\acknowledgments
\emph{Acknowledgments} The authors thank the referee for their helpful comments.  This research has made use of the SIMBAD database, operated at CDS, Strasbourg, France.  This research has made use of NASA's Astrophysics Data System.  This publication makes use of data products from the Two Micron All Sky Survey, which is a joint project of the University of Massachusetts and the Infrared Processing and Analysis Center/California Institute of Technology, funded by the National Aeronautics and Space Administration and the National Science Foundation.  This research has made use of the NASA/ IPAC Infrared Science Archive, which is operated by the Jet Propulsion Laboratory, California Institute of Technology, under contract with the National Aeronautics and Space Administration.  This research was supported by an appointment to the NASA Postdoctoral Program at the Ames Research Center, administered by the Oak Ridge Associated Universities through a contract with NASA.

\clearpage

%% This is Figure 1 
\begin{figure}
\plotone{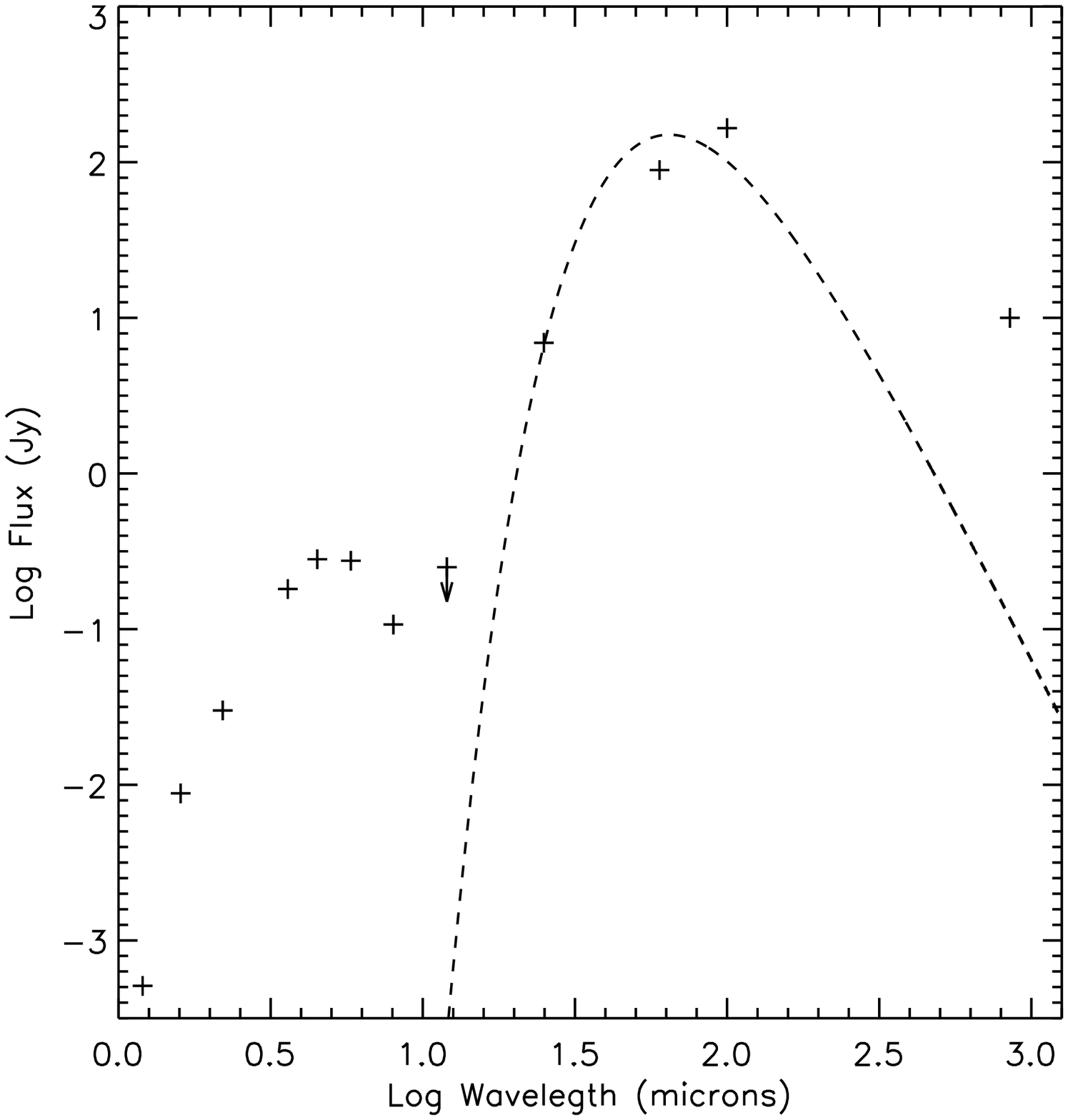}
\caption{The SED of L483, combining data from 2MASS, Spitzer, IRAS and an 850~$\mu$m observation from \citet{Ful2000}. There is significant emission at wavelengths shortward of 10~$\mu$m, and the SED is not well fit by a single temperature blackbody, which is uncharacteristic of a Class 0 YSO.  The dotted line is a blackbody with a temperature of 45~K, which was temperature of the best-fit blackbody function to the IRAS data.  We did not fit a separate blackbody function to the flux values from 1.2~$\mu$m to 8.0~$\mu$m since fluxes in that wavelength range are strongly affected by extinction and scattered light.  \label{fig3}}
\end{figure}

%% This is Figure 2 
 \begin{figure}
 \plotone{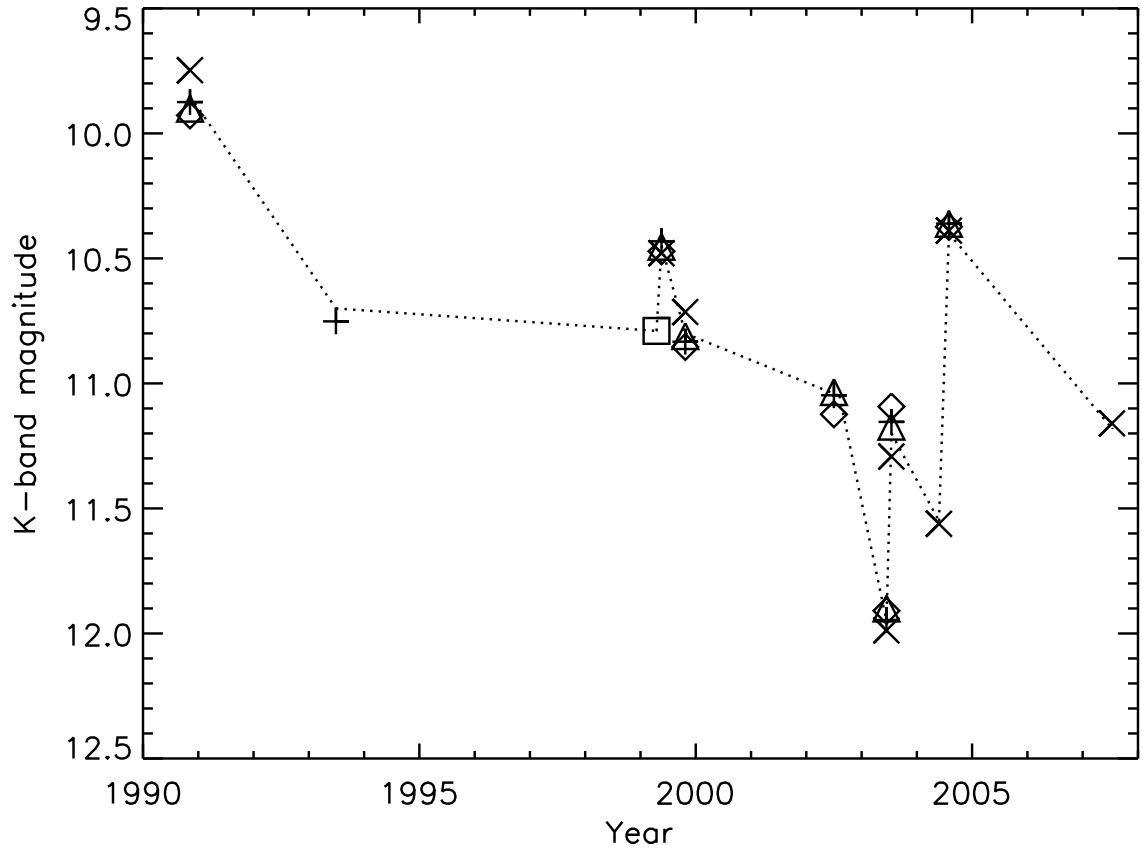}
 \caption{The variability of the integrated flux of the L483 infrared nebula at K-band. The plus, diamond, triangle, and 'x' symbols represent photometry of the nebula relative to four different field stars.  The square is the K-band magnitude measured by 2MASS.  The dotted line passes through the average value at each epoch.   \label{fig4}}
 \end{figure}

%% This is Figure 3
\begin{figure}
\plotone{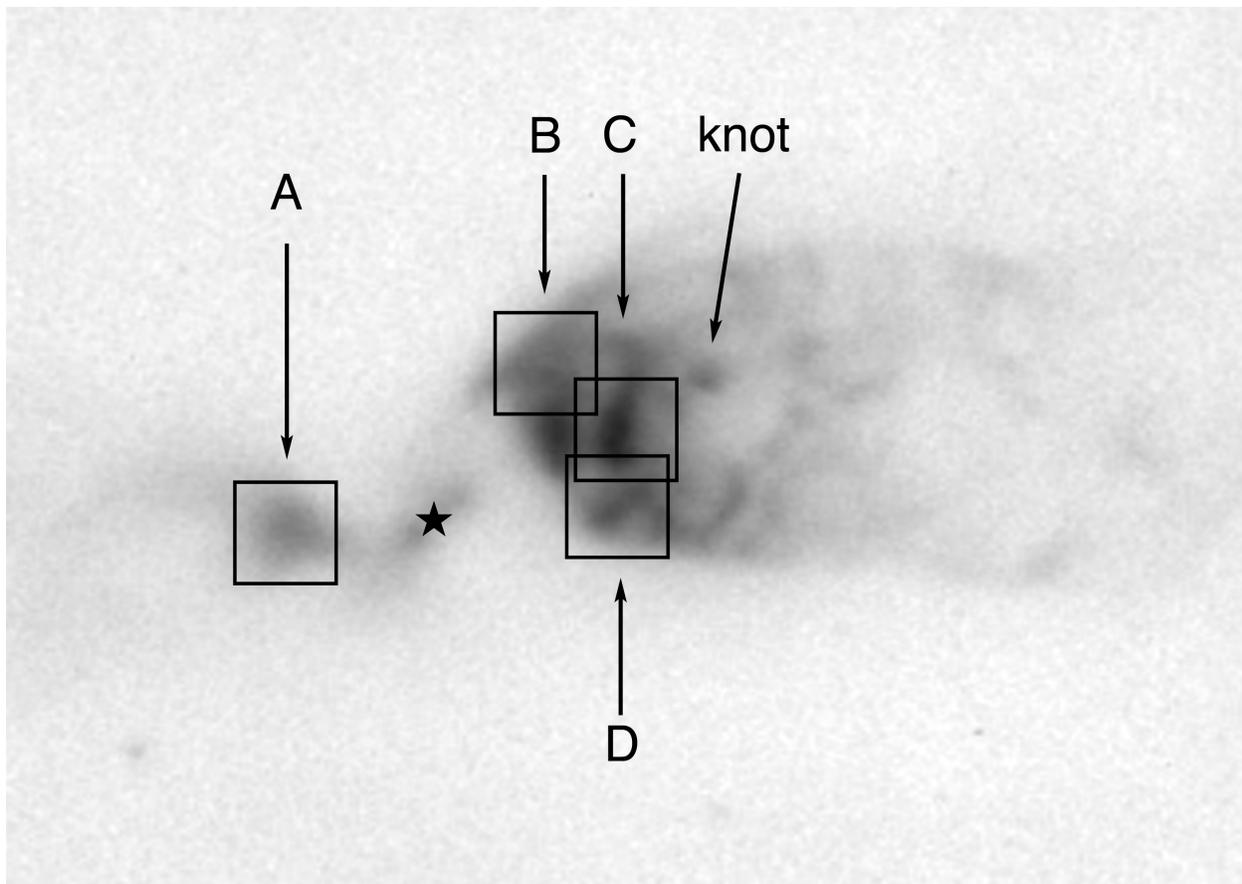}
\caption{The master image with labels for four regions as well as a knot.  Changes in the brightness of these regions are discussed in section 3.2.  The location of the 24~$\mu$m source observed by Spitzer is marked with the star symbol. \label{fig3}}
\end{figure}

%% This is Figure 4.1
\begin{figure}
\plotone{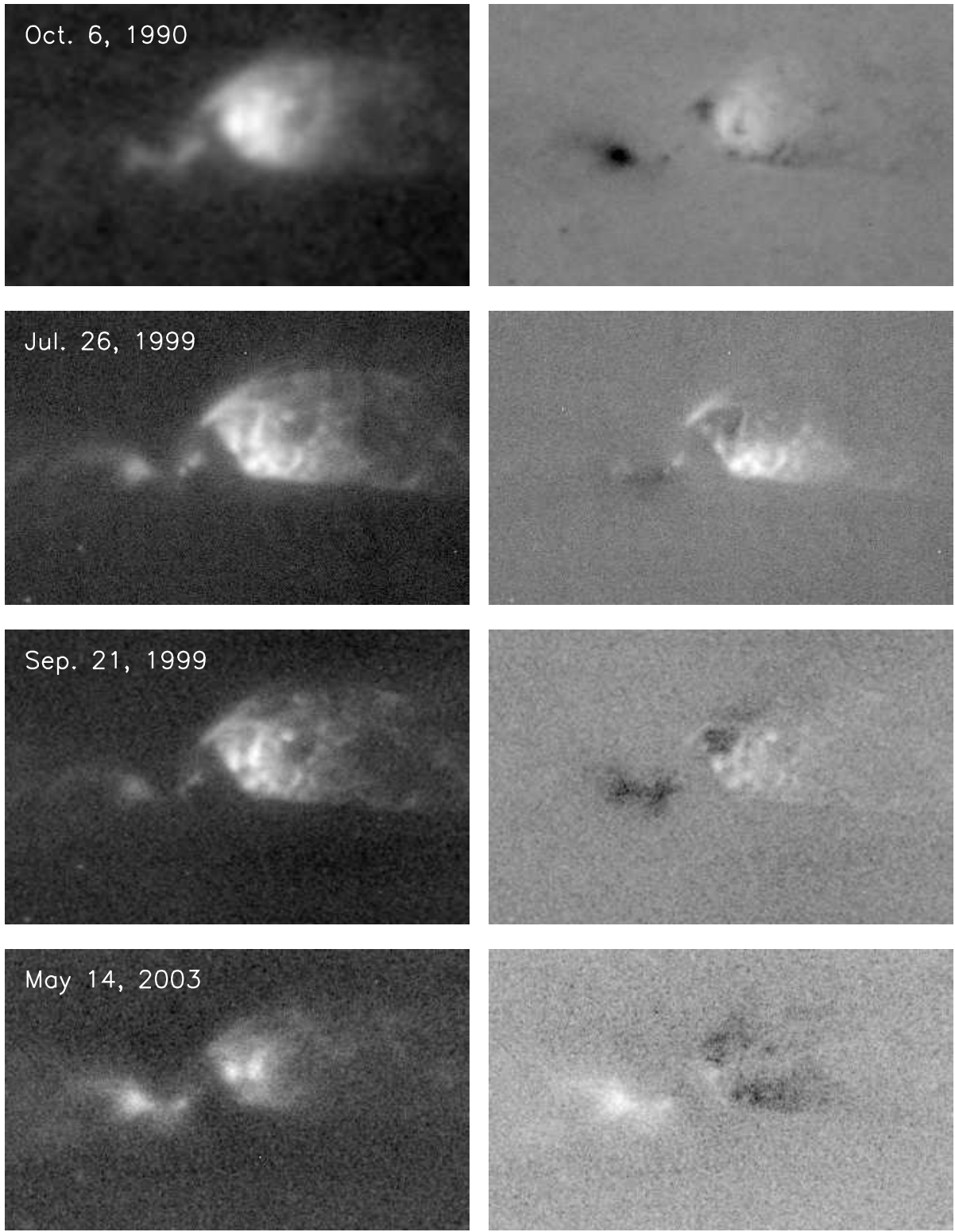}
\caption{ 30\arcsec~ by 20\arcsec~ views of L483 at K or K$'$ at 8 epochs taken over a span of 17 years. The grayscale was chosen to best show the morphology of the nebula, and does not represent the actual surface brightness of the nebula.  The left column shows the images of the nebula whereas the right column shows the difference between the image on the left and the master image.  The image taken in July 2002 is not deep enough to clearly show morphological features. \label{fig3}}
\end{figure}

%% This is Figure 4.2
\begin{figure}
\plotone{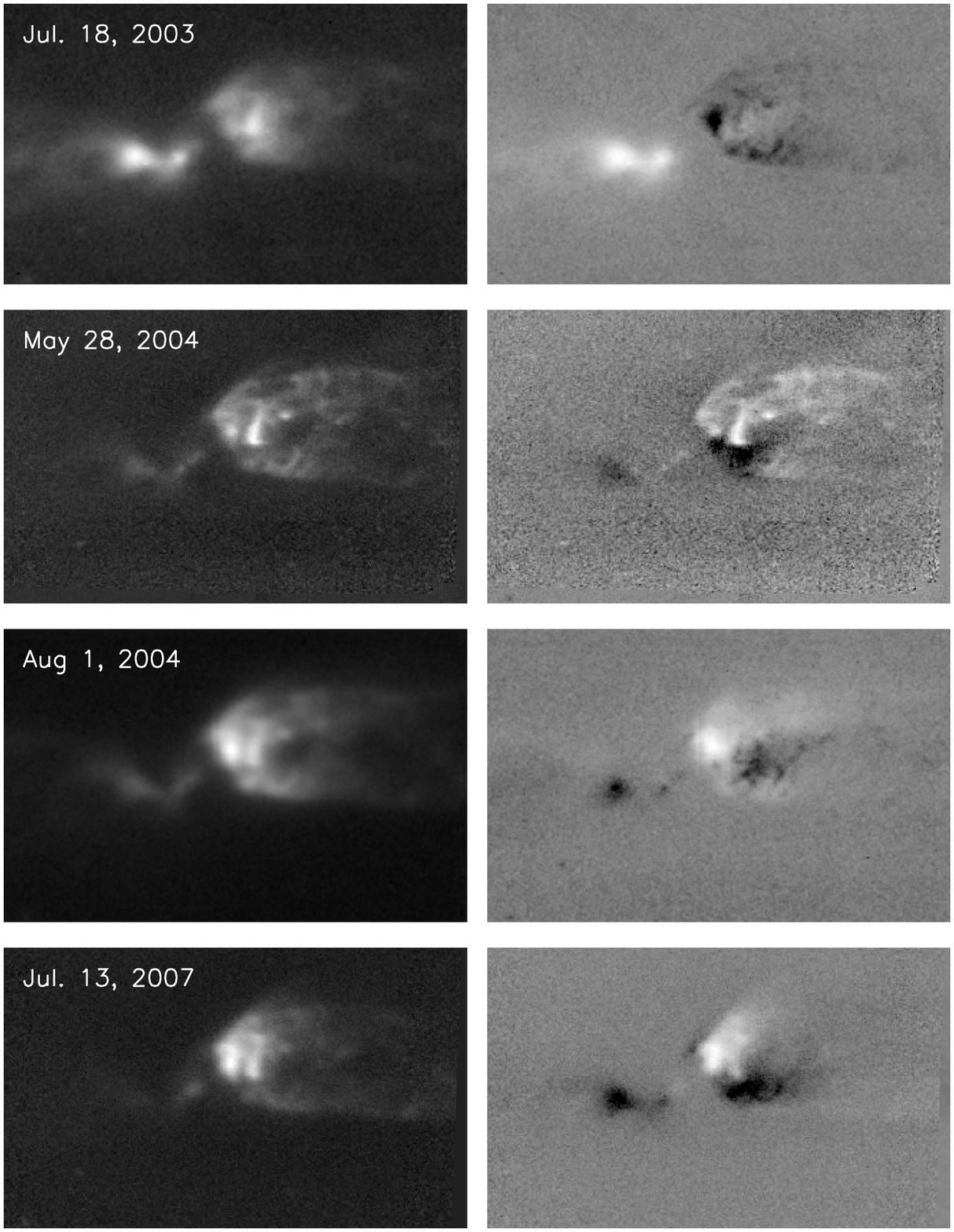}
\addtocounter{figure}{-1}
\caption{ Figure 3 continued.  \label{fig3}}
\end{figure}

%% This is Figure 5
 \begin{figure}
 \plotone{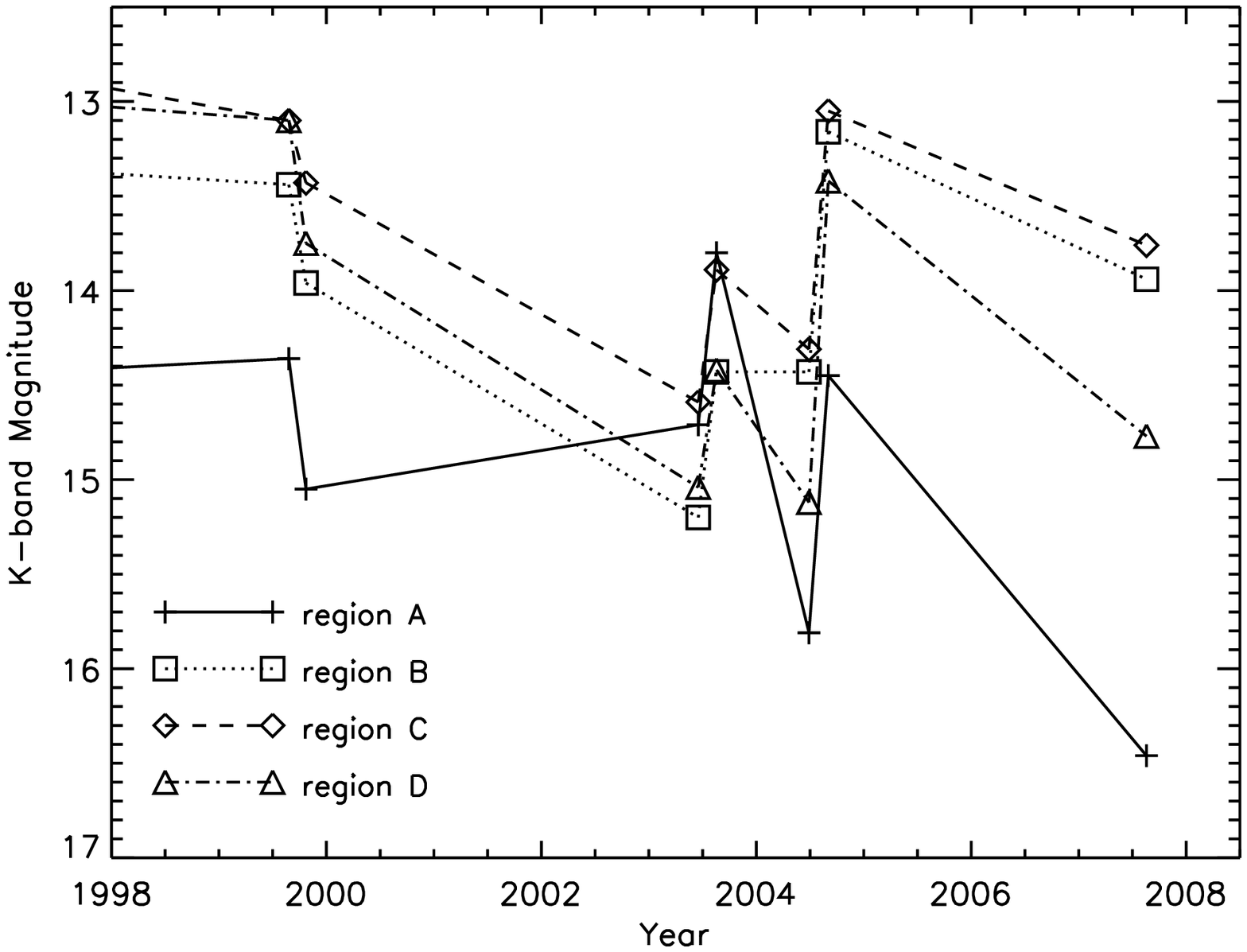}
 \caption{K-band photometry of four regions within the nebula.  Data from October 1990 is to the left of the plot for clarity.  Generally, different regions of the nebula brighten and fade together, occasionally by different amounts.  However, there are notable exceptions, particularly between 2003 and 2004.  The image taken in July 2002 is not deep enough to allow us to measure the photometry in different regions of the nebula.  \label{fig4}}
 \end{figure}

%% Here we use \plottwo to present two versions of the same figure,
%% one in black and white for print the other in RGB color
%% for online presentation. Note that the caption indicates
%% that a color version of the figure will be available online.
%%

%% This is Figure 6
\begin{figure}
\plotone{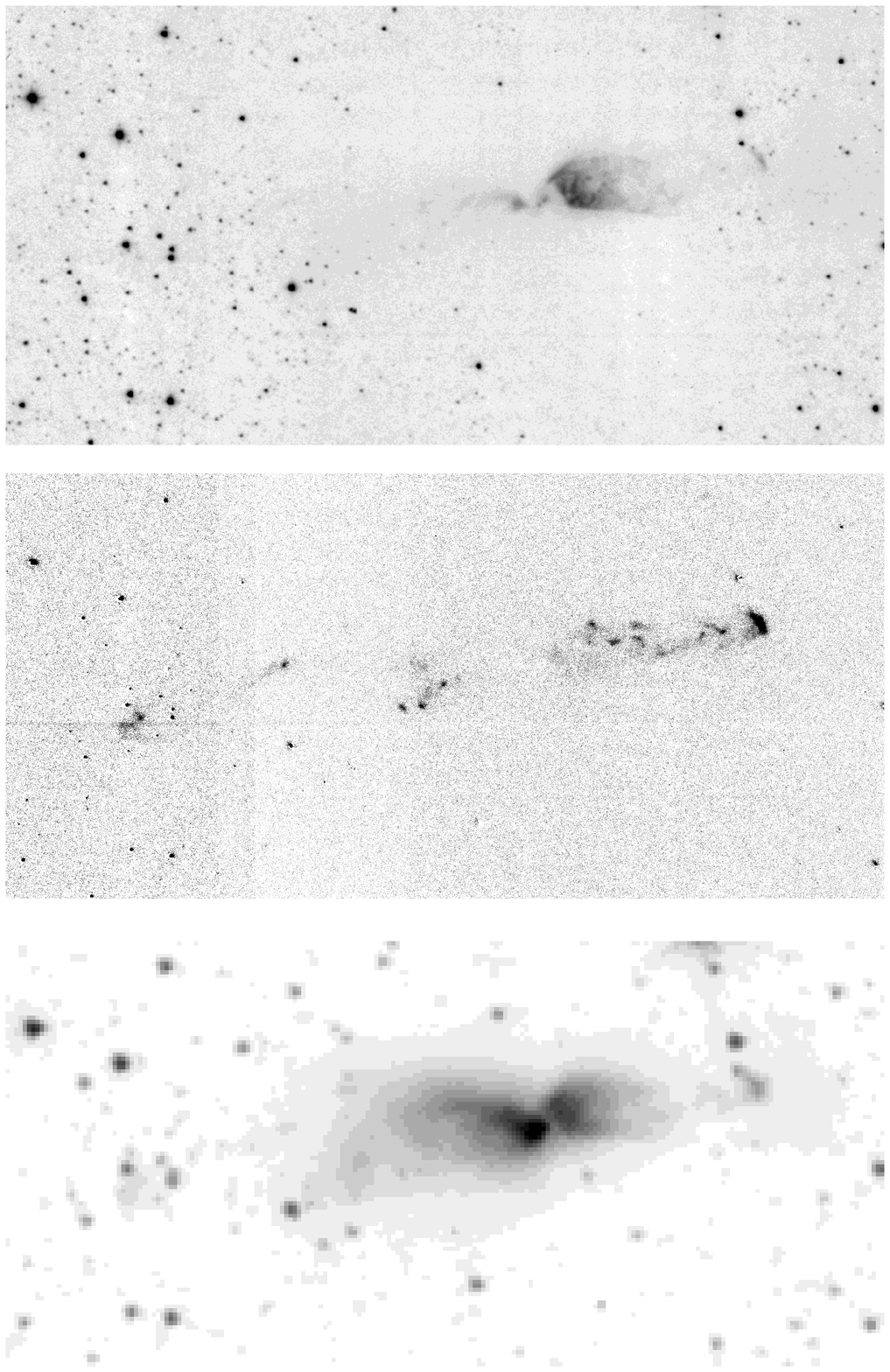}
\caption{120\arcsec~ by 60\arcsec~ views of L483 with North up and East to the left.  The top frame is a K-band image from July 1999.  The source lies behind the small, faint nebula between the main nebula and the fainter nebula to the East.  Note the dearth of field stars within $\sim45$\arcsec~ of the source.  The middle frame is our H$_{2}$ image with the K-band image from the same night subtracted out.  The bright shock to the west is also visible in the K-band image.  The H$_{2}$ shocks extend farther to the east than the west from the location of the source.  The bottom frame is a Spitzer IRAC 8.0~$\mu$m image.  Both lobes of its outflow cavity are now clearly visible.  The bright source near the center of the nebula has over twice the FWHM of field stars, and thus is likley scattered light from the central star, which remains obscured. \label{fig1}}
\end{figure}

\clearpage
%% L483 observations table

\begin{deluxetable}{lccccccccccccc}
\tabletypesize{\scriptsize}
\tablecaption{Observations}
\tablewidth{0pt}
\tablecolumns{3}
\tablehead{
\colhead{Date (UT)} & \colhead{Observer} &
\colhead{Filter} & \colhead{FWHM ('')} & \colhead{Instrument}
}
\startdata

%%  Date          Observer       Filter            FWHM    Instr.
Oct.  6, 1990  & Hodapp    &  K$'$               &  0.92 &  NICMOS3  \\
June 27, 1993  & Fuller    &  K                  & 2.68 &  IRIM  \\
Apr. 15, 1999  & 2MASS     &  K                  & 2.31 &  2MASS  \\
July 25, 1999  & Hodapp    &  K                  & 0.52 &  QUIRC  \\
July 25, 1999  & Hodapp    &  H$_{2}$ v=1-0 S(1) & 0.64 &  QUIRC  \\
July 26, 1999  & Hodapp    &  H                  & 0.60 &  QUIRC  \\
Sep. 21, 1999  & Hodapp    &  K$'$              & 0.55 &  QUIRC  \\
July  3, 2002  & Hodapp    &  K$'$              & 1.05 &  QUIRC  \\
May  14, 2003  & Connelley &  K                  & 1.15 & QUIRC   \\
July 18, 2003  & Connelley &  K                  & 1.38 & QUIRC   \\
May  28, 2004  & Connelley &  K                  & 0.70 & SpeX   \\
Aug.  1, 2004  & Connelley &  K                  & 1.26 & QUIRC   \\
July 13, 2007  & Connelley &  K                  & 1.02 & SpeX   \\

\enddata

\end{deluxetable}

\clearpage


\begin{thebibliography}{}
\bibitem[()]{} 

\bibitem[Andr\'{e} et al.(2000)]{And2000}Andr\'{e}, P., Ward-Thompson, D., \& Barsony, M., 2000, Protostars and Planets IV, eds. V. Mannings, A. Boss, \& S. Russell, Univ. of Arizona Press, Tucson, p.59

\bibitem[Barnard(1895)]{Bar1895} Barnard, E., 1895, MNRAS, 55, 442 

\bibitem[Brice\~{n}o et al.(2004)]{Bri2004} Brice\~{n}o, C., Vivas, A., Hern\'{a}ndez, J., Calvet, N., Hartmann, L., Megeath, T., Berlind, P., Calkins, M., \& Hoyer, S., 2004, \apj, 606, L123  

\bibitem[Buckle et al.(1999)]{Buc1999} Buckle, J., Hatchell, J., \& Fuller, G., 1999, \aap,  348, 584

\bibitem[Cohen et al.(1977)]{Coh1977} Cohen, M., Kuhi, L., \& Harlan, E., 1977, ApJ, 215L, 127  

%% \bibitem[Cohen et al.(1981)]{Coh1981} Cohen, M., Kuhi, L., Harlan, \& E., Spinrad, H., 1981, \apj, 245, 920  

\bibitem[Connelley et al.(2007)]{Con2007} Connelley, M., Reipurth, B., \& Tokunaga, A., 2007, \aj, 133, 1528 %% the nebula paper

\bibitem[Dame \& Thaddeus(1985)]{Dam1985} Dame, T., \& Thaddeus, P., 1985, \apj, 297, 751

\bibitem[Evans et al.(2003)]{Eva2003} Evans, N., Allen, L., Blake, G., Boogert, A., Bourke, T., Harvey, P., \& Kessler, J., 2003, PASP, 115, 965

\bibitem[Fedele et al.(2007)]{Fed2007} Fedele, D., van den Ancker, M., Petr-Gotzens, M., Ageorges, N., \& Rafanelli, P., 2007, \aap, 472, 199

\bibitem[Felli et al.(1992)]{Fel1992} Felli, M., Palagi, F., \& Tofani, G., 1992, \aap, 255, 293

\bibitem[Fuller et al.(1995)]{Ful1995} Fuller, G., Lada, E., Masson, C., \& Myers, P., 1995, \apj, 453, 754 

\bibitem[Fuller \& Wootten(2000)]{Ful2000} Fuller, G., \& Wootten, A., 2000, \apj, 534, 854 

\bibitem[Goodman et al.(1993)]{Goo1993} Goodman, A., Benson, P., Fuller, G., \& Myers, P., 1993, \apj, 453, 754

\bibitem[Gyul$'$budagyan \& Magakyan(1977)]{Gyu1977} Gyul$'$budagyan, A., \& Magakyan, T., 1977, Soviet Astronomy Letters, 3, 58

\bibitem[Hind(1864)]{Hin1864} Hind, J., 1864, MNRAS, 24, 65 

\bibitem[Hodapp(1994)]{Hod1994} Hodapp, K., 1994, ApJS, 94, 615 

\bibitem[Hodapp et al.(1996)]{Hod1996} Hodapp, K., et al. 1996, New Astronomy, 1, 177

\bibitem[Hodapp et al.(2004)]{Hod2004} Hodapp, K., Walker, C., Reipurth, B., Wood, K., Bally, J., Whitney, B., \& Connelley, M., 2004, \apj, 601, L79

\bibitem[Hodapp \& Bressert (2008)]{Hod2008} Hodapp, K., \& Bressert, E., 2008, in press

\bibitem[Hubble(1916)]{Hub1916} Hubble, E., 1916, \apj, 44, 190

\bibitem[Knox Shaw(1916)]{Kno1916} Knox Shaw, H., 1916, MNRAS, 76, 646

%% \bibitem[Koresko et al.(1999)]{Kor1999} Koresko, C., Blake, G., Brown, M., Sargent, A., \& Koerner, D., 1999, \apj, 525L, 49 

\bibitem[Lada(1991)]{Lad1991} Lada, C., 1991, in The Physics of Star Formation and Early Stellar Evolution, eds. C. J. Lada \& N. D. Kylafis, Kluwer Academic Publishers, p.329

\bibitem[Lampland(1926)]{Lam1926} Lampland, C., 1926, Popular Astronomy, 34, 621 

\bibitem[Parker(1988)]{Par1988} Parker, N., 1988, MNRAS, 235, 139 

\bibitem[Pezzuto et al.(2002)]{Pez2002}Pezzuto, S., et al., 2002, \mnras, 330, 1034

\bibitem[Pontoppidan \& Dullemond (2005)]{Pon2005} Pontoppidan, K., \& Dullemond, C., 2005, \aap, 435, 595

\bibitem[Rayner et al.(2003)]{Ray2003} Rayner, J., Toomey, D., Onaka, P., Denault, A., Stahlberger, W., Vacca, W., Cushing, M., \& Wang, S., 2003, PASP, 15, 362


\bibitem[Reipurth \& Aspin(2004)]{Rei2004} Reipurth, B., \& Aspin, C., 2004, \apj, 606, L119

\bibitem[Simons \& Tokunaga(2002)]{Sim2002} Simons, D., \& Tokunaga, A., 2002, PASP, 114, 169

\bibitem[Skrutskie et al.(2006)]{Skr2006} Skrutskie, M., et al. 2006, AJ, 131, 1163

\bibitem[Tafalla et al.(2000)]{Taf2000} Tafalla, M., Myers, P., Mardones, D., \& Bachiller, R., 2000, \aap, 359, 967 

\bibitem[Tokunaga \& Simons(2002)]{Tok2002} Tokunaga, A., \& Simons, D., 2002, PASP, 114, 180

\bibitem[Wainscoat \& Cowie(1992)]{Wai1992} Wainscoat, R., \& Cowie, L., 1992, \aj, 103, 332 

\end{thebibliography}
\end{document}